\documentclass[11pt,draftclsnofoot,conference,onecolumn]{IEEEtran}

\usepackage{amssymb}    
\usepackage{color}      

\usepackage{cite}       

\usepackage{graphicx}   

\usepackage{url}        
\newtheorem{remark}{Remark}
\newtheorem{condition}{Condition}

\newcommand{\n}{\mathcal{N}}
\newcommand{\z}{\mathbb{Z}}
\newcommand{\f}{\mathbb{F}}
\newcommand{\A}{\overline{A}}
\newcommand{\B}{\overline{B}}
\newcommand{\C}{\overline{C}}
\newcommand{\I}{\overline{I}}
\newcommand{\Ba}{\overline{B_1}}
\newcommand{\Bc}{\overline{B_3}}
\newcommand{\Bd}{\overline{B_4}}

\begin{document}

\title{Violating the Ingleton Inequality with Finite Groups}

\author{\authorblockN{Wei Mao, Babak Hassibi}
\authorblockA{Department of Electrical Engineering\\
California Institute of Technology, Pasadena, CA 91125\\
Email: \{wmao, hassibi\}@caltech.edu}}


%


\maketitle

\begin{abstract}
It is well known that there is a one-to-one correspondence between the
entropy vector of a collection of $n$ random variables and a certain
group-characterizable vector obtained from a finite group and $n$ of
its subgroups \cite{Entropy_Group}. However, if one restricts
attention to abelian groups then not all entropy vectors can be
obtained. This is an explanation for the fact shown by Dougherty et al
\cite{zeger} that linear network codes cannot achieve capacity in
general network coding problems (since linear network codes form an
abelian group). All abelian group-characterizable vectors, and by fiat
all entropy vectors generated by linear network codes, satisfy a
linear inequality called the Ingleton inequality. In this paper, we
study the problem of finding nonabelian finite groups that yield
characterizable vectors which violate the Ingleton inequality. Using
a refined computer search, we find the symmetric group $S_5$ to be the
smallest group that violates the Ingleton inequality. Careful study of
the structure of this group, and its subgroups, reveals that it belongs to
the Ingleton-violating family $PGL(2,p)$ with primes $p \geq 5$, i.e.,
the projective group of $2\times 2$ nonsingular matrices with entries
in $\f_p$. This family of groups is therefore a good candidate for
constructing network codes more powerful than linear network codes.
\end{abstract}


%
\IEEEpeerreviewmaketitle

\section{Introduction}
\label{section:intro}
Let $\n = \{1,2,\dots,n\}$, and let $X_1,X_2,\dots,X_n$ be $n$ jointly
distributed discrete random variables. For any nonempty set $\alpha
\subseteq \n$, let $X_\alpha$ denote the collection of random
variables $\{X_i:i\in\alpha\}$, with joint entropy $h_\alpha\triangleq
H(X_\alpha) = H(X_i;\,i\in\alpha)$. We call the ordered real $(2^n-1)$-tuple
$(h_\alpha:\emptyset \neq \alpha \subseteq \n) \in \mathbb{R}^{2^n-1}$ an
entropy vector. The set of all entropy vectors
derived from $n$ jointly distributed discrete random variables is
denoted by $\Gamma_n^*$. It is not too difficult to show that the
closure of this set, i.e., $\overline{\Gamma_n^*}$, is a {\em convex
  cone}.

The set $\overline{\Gamma_n^*}$ figures prominently in information theory
since it describes the possible values that the joint entropies of a
collection of $n$ discrete random variables can obtain. From a
practical point of view, it is of importance since it can be shown
that the capacity region of any arbitrary multi-source multi-sink {\em
  wired} network, whose graph is acyclic and whose links are discrete
memoryless channels, can be obtained by optimizing a linear function
of the entropy vector over the convex cone $\overline{\Gamma_n^*}$ and a set
of linear constraints (defined by the network)
\cite{yanyeungzhen07,hash07}. Despite this importance, the entropy
region $\overline{\Gamma_n^*}$ is only known for $n=2,3$ random variables
and remains unknown for $n\geq 4$ random variables. Nonetheless, there
are important connections known between $\overline{\Gamma_n^*}$ and
matroid theory (since entropy is a submodular function and therefore somehow
defines a matroid) \cite{dofrze07}, determinantal inequalities
(through the connection with Gaussian random variables) \cite{shha08},
and quasi-uniform arrays \cite{chan01}. However, perhaps most
intriguing is the connection to finite groups which we briefly
elaborate below.

\subsection{Groups and Entropy}
\label{subsec:group}

Let $G$ be a finite group, and let $G_1,G_2,\dots,G_n$ be $n$ of its
subgroups. For any nonempty set $\alpha \subseteq \n$, the group $G_\alpha
\triangleq \cap_{i\in\alpha}G_i$ is a subgroup of $G$. Let $|K|$ be the order
(cardinality) of a group $K$, and define $g_\alpha\triangleq
\log\frac{|G|}{|G_\alpha|}$. We call the ordered real
$(2^n-1)$-tuple $(g_\alpha: \emptyset\neq \alpha \subseteq \n) \in
\mathbb{R}^{2^n-1}$ a (finite) group characterizable vector. Let
$\Upsilon_n$ be the set of all group characterizable vectors derived
from $n$ subgroups of a finite group.

The major result shown by Chan and Yeung in \cite{Entropy_Group} is
that $\overline{\Gamma_n^*} =
\overline{\mathrm{cone}(\Upsilon_n)}$, i.e., the closure of
$\Gamma_n^*$ is the same as the closure of the cone generated by
$\Upsilon_n$. In other words, every group characterizable vector is
an entropy vector, whereas every entropy vector is arbitrarily
close to a scaled version of some group characterizable vector.

To show that every group characterizable vector is an entropy vector
\cite{Entropy_Group} gives the following construction. Let $\Lambda$ be
be a random variable uniformly distributed on the elements of $G$. Now
for $i=1,\ldots,n$ define $X_i = \Lambda G_i$ (the left coset of
$\Lambda$ in $G$ w.r.t. the subgroup $G_i$). Then a simple calculation
shows that $h_\alpha = \log\frac{|G|}{|G_\alpha|} = g_\alpha$,
implying that every group-characterizable vector is an entropy
vector. Showing the other direction, i.e., that every entropy vector
is arbitrarily close to a scaled version of a group-characterizable
vector is more tricky (the interested reader may consult
\cite{Entropy_Group} for the details). Here we shall briefly describe
the intuition.

Consider a random variable $X_1$ with alphabet size $N$ and probability
mass function $\{p_i,i=1,\ldots,N\}$. Now if we make $T$ copies of
this random variable to make sequences of length $T$, the entropy of
$X_1$ is roughly equal to the logarithm of the number of typical
sequences. These are
sequences where $X_1$ takes its first value roughly $Tp_1$ times, its
second value roughly $Tp_2$ times and so on. Therefore assuming that $T$ is
large enough so that the $Tp_i$ are close to integers (otherwise, we
have to round things) we may roughly write
\[ H(X_1) \approx \frac{1}{T}\log\left(\begin{array}{ccccc} & & T & &
  \\ Tp_1 & Tp_2 & \ldots & Tp_{N-1} & Tp_N \end{array} \right), \]
where the argument inside the $\log$ is the usual multinomial
coefficient. Written in terms of factorials this is
\begin{equation}
H(X_1) \approx \frac{1}{T}\log\frac{T!}{(Tp_1)!(Tp_2)!\ldots
  (Tp_N)!}.
\label{firstent}
\end{equation}
If we consider the group $G$ to be the symmetric group $S_T$, i.e.,
the group of permutations among $T$ objects, then clearly $|G| =
T!$. Now partition the $T$ objects into $N$ sets each with $Tp_1$ to
$Tp_N$ elements, respectively, and define the group $G_1$ to be the
subgroup of $S_T$ that permutes these objects {\em while respecting the
  partition}. Clearly, $|G_1| = (Tp_1)!(Tp_2)!\ldots (Tp_N)!$, which
is the denominator in (\ref{firstent}). Thus, $H(X_1) \approx
\frac{1}{T}\log\frac{|G|}{|G_1|}$, so that the entropy $h_{\{1\}}$ is a scaled
version of the group-characterizable $g_{\{1\}}$. This argument can be made
more precise and can be extended to $n$ random variables---see
\cite{Entropy_Group} for the details. We note, in passing, that this
construction often needs $T$ to be very large, so that the group $G$
and the subgroups $G_i$ are huge.

\subsection{The Ingleton Inequality}

As mentioned earlier, entropy satisfies submodularity and therefore, with some care,
defines a matroid. Matroids are defined by a ground set and a rank
function, defined over subsets of the ground set, that satisfies
submodularity. They were defined in a way to extend the notion of a
collection of vectors (in some vector space) along with the usual
definition of the rank. A matroid is called
{\em representable} if its ground set can be represented as a
collection of vectors (defined over some finite field) along with the
usual rank function. Determining whether a matroid is representable or
not is, in general, an open problem.

Let $n = 4$, $\n =\{1,2,3,4\}$. In 1971 Ingleton showed that the rank
function $x_{\{\cdot\}}$ of any representable matriod must satisfy the
inequality \cite{ingleton}
\begin{equation}\label{equation:Ingletonh}
x_{12}+x_{13}+x_{14}+x_{23}+x_{24} \geq
x_1+x_2+x_{34}+x_{123}+x_{124}
\end{equation}
where for simplicity we write $x_{ij}$ and $x_{ijk}$ for
$x_{\{i,j\}}$ and $x_{\{i,j,k\}}$, respectively. However, it turns out
that there are entropy vectors that violate the Ingleton inequality
\cite{matus95}, so that entropy is generally not a representable
matroid. Using non-representable matroids, \cite{zeger} constructs
network coding problems that cannot be solved by linear network codes
(since linear network codes are, by definition, representable).

As $\overline{\Gamma_n^*} = \overline{\mathrm{cone}(\Upsilon_n)}$, we
know there must exist finite groups, and corresponding subgroups, such
that their induced group-characterizable vectors violate the Ingleton
inequality. In \cite{Group_Entropy} it was shown that abelian groups
cannot violate the Ingleton inequality, thereby giving an alternative
proof as to why linear network codes cannot achieve capacity on
arbitrary networks---they form an abelian group. So we need to focus
on non-abelian groups and their connections to nonlinear codes.

Finally, we remark that, in the context of finite groups, the Ingleton
inequality can be rewritten as
\begin{equation}\label{equation:Ingletong}
|G_1||G_2||G_{34}||G_{123}||G_{124}| \geq
|G_{12}||G_{13}||G_{14}||G_{23}||G_{24}|
\end{equation}

\subsection{Discussion}

Since we know of distributions whose entropy vector violates the
Ingleton inequality, we can, in principle, construct finite groups
whose group-characterizable vectors violate Ingleton. Two such
distributions are Example 1 in \cite{exVioIngleton}, where the
underlying distribution is uniform over 7 points and the random
variables correspond to different partitions of these seven points,
and the example on page 1445 of \cite{ZY98}, constructed from finite
projective geometry and where the underlying distribution is uniform
over $12\times 13 = 156$ points. Unfortunately, constructing groups
and subgroups for these distributions using the recipe of section
\ref{subsec:group} results in $T = 29\times 7 = 203$ and $T = 23\times
156 = 3588$, which results in groups of size $203!$ and $3588!$, which are too
huge to give us any insight whatsoever.

These discussions lead us to the following questions.

\begin{enumerate}
\item Could the connection between entropy and groups be a red
  herring? Are the interesting groups too large to give any insight
  into the problem (e.g., the conditions for the Ingleton inequality
  to be violated)?
\item What is the smallest group with subgroups that violates the
  Ingleton inequality? Does it have any special structure?
\item Can one construct network codes from such Ingleton-violating
  groups?
\end{enumerate}

In this paper we address the first two questions. We
identify the smallest group that violates the Ingleton inequality---it
is the symmetric group $S_5$, with 120 elements. Through a thorough
investigation of the structure of its subgroups we conclude that it
belongs to the family of groups $PGL(2,p)$, with $p$ a prime greater
than or equal to 5. ($PGL(2,5)$ is isomorphic to
$S_5$.)\footnote{$GL(2,p)$ is the general linear group of invertible
  $2\times 2$ matrices with entries in $\f_p$. $PGL(2,p)$ is the {\em
    projective} general linear group, where proportional matrices in $GL(2,p)$
  are all mapped to the same element.} We therefore believe that the
connection to groups is not a red herring and that there may be some
benefit to it.

The explicit nature of $PGL(2,p)$ may lend itself to
effective network codes. We only mention that non-abelian groups allow
for much more flexibility in the design of codes. For example, if the
incoming messages to a node in the network, $a$ and $b$, say, are
elements from a nonabelian group then the operations $a^2b$, $aba$,
$ba^2$, say, can potentially all correspond to different elements in
the group, whereas in the abelian case they all coincide with
$a^2b$. Therefore nodes in a network will have much more choices in
terms of what to transmit on their outgoing edges---and this should,
ostensibly, be what allows one to achieve capacity. The drawback is, of
course, that decoding becomes more complicated than solving a system
of linear equations.

We shall not say anymore about codes. What we will do in the remainder
of the paper is to describe how we found the smallest Ingleton-violating
group and how we uncovered its structure. This required the
identification of conditions beyond being abelian that force a group to
respect Ingleton. It also required a deep study of the 120 element
group that we found via computer search. We now present the details.

\section{Notation}

We use the following abstract algebra notations throughout this
paper:\\
\begin{tabular}{lcl}
$|G|$ &:& the order of group $G$.\\
$G \cong H$ &:& the group $G$ is isomorphic to the group $H$.\\
$H \leq G$, $H < G$ &:& $H$ is a subgroup of $G$, and a proper subgroup of $G$.\\
$H \trianglelefteq G$ &:& $H$ is a normal subgroup of $G$.\\
$G/H$ &:& the set of all left cosets of subgroup $H$ in $G$. When $H
\trianglelefteq G$, $G/H$ is a group. \\&&(Factor or quotient group)\\
$|g|$ &:& the order of element $g$ = smallest positive integer $m$ s.t. $g^m = 1$.\\
$x^g$ &:& the conjugate of element $x$ by element $g$ in $G$:
$x^g=g^{-1}xg$. \\&&(No confusion with the powers of $x$ as $g$ is an element of $G$.)\\
$X^g$ &:& the conjugate of subset $X$ by element $g$ in $G$:
$X^g=\{x^g:x\in X\}$.\\
$HK$ &:& the ``set product'' of $H,K\subseteq G$: $HK = \{hk:h\in H, k\in K\}$.\\
$H\rtimes K$ &:& the semidirect product of groups $H$ and $K$.\\
$\langle g_1,\dots,g_m \rangle$, $\langle S \rangle$ &:& the group generated by the elements $g_1,\dots,g_m$, and by the set $S$.\\
$G=\langle S|R \rangle$ &:& $\langle S|R \rangle$ is a presentation of $G$.
$S$ is a set of generators of $G$, while $R$ is a set\\&& of relations $G$ should satisfy.\\
$1$ &:& the natural number ``1'', identity element of a group, or
the trivial group.\\&&The meaning should be clear in different
contexts with no confusion.\\
$\z_n$ &:& the integers modulo $n$ $\cong$ the cyclic group of order $n$.\\
$S_n$ &:& the symmetric group of degree $n$ = all permutations on $n$ points.\\
\end{tabular}\\
\begin{tabular}{lcl}
$D_{2n}$ &:& the dihedral group of order $2n$.\\
$\f_q$ &:& the finite field of $q$ elements.\\
$\z_n^\times$, $\f_q^\times$ &:& the multiplicative group of units of $\z_n$, and of $\f_q$. $\f_q^\times$ = all nonzero elements of $\f_q$.\\
$GL(n,q)$ &:& the general linear group of all invertible $n \times n$ matrices with entries from $\f_q$.\\
&&The identity element for $GL(n,q)$ is usually denoted by $I$ = identity matrix.\\
$PGL(n,q)$ &:& the projective general linear group = $GL(n,q)/V$, where $V$ = all nonzero scalar\\
&& matrices = $\{\alpha I:\alpha \in\f_q^\times\}$.\\
\end{tabular}

\section{Computer Search and Some Negative Conditions}
Designing a small admissible structure for the group $G$ and its
subgroups without an existing Ingleton-violating instance is very
difficult, so we use computer programs to search for a small
instance. We use the GAP system\cite{GAP4} to search its ``Small
Group'' library, which contains all finite groups of order less than
or equal to 2000 except 1024. We pick a group in this library, find
all its subgroups, then test Ingleton inequality for all
4-combinations of these subgroups. This is a tremendous task, as
there are already more than 1000 groups of order less than or equal
to 100, each of which might have hundreds of subgroups (some even
have more than 1000).

It was therefore extremely critical to prune our search. In fact, we
used the following ``negative conditions'',
each of which guarantees that Ingleton is never violated.

\begin{condition}\label{condition:abelian}
$G$ is abelian.\cite{Group_Entropy}
\end{condition}

\begin{condition}\label{condition:normal}
$G_i \trianglelefteq G$, $\forall i$.\cite{Ingleton_Hom}
\end{condition}

\begin{condition}\label{condition:g1g2}
$G_1G_2 = G_2G_1$, or equivalently $G_1G_2 \leq G$.
\end{condition}

\begin{proof}
(sketch) Construct random variables $X_i$'s from uniformly
distributed $\Lambda$ on $G$ as in Section~\ref{subsec:group}. As
$G_{1;2} \triangleq G_1G_2 \leq G$, we can similarly construct
random variable $X_{1;2} = \Lambda G_{1;2}$. Note that $|G_{1;2}|=
|G_1||G_2|/|G_{12}|$, $H(X_{1;2}|X_1) = H(X_{1;2}|X_2) = 0$ as
$G_1,G_2\leq G_{1;2}$. Similar to the proof of Condition
\ref{condition:normal} in \cite{Ingleton_Hom}, we use the following
information inequality in \cite{techIneq}:
\begin{displaymath}
2H(E|A) + 2H(E|B) + I(A;B|C) + I(A;B|D) + I(C;D) \geq H(E).
\end{displaymath}
Plugging in $A = X_1$, $B = X_2$, $C = X_3$, $D = X_4$ and $E =
X_{1;2}$ one can easily deduce Ingleton inequality.
\end{proof}

\begin{remark}
Condition \ref{condition:normal} subsumes Condition
\ref{condition:abelian}, while Condition \ref{condition:g1g2}
subsumes Condition \ref{condition:normal}.
\end{remark}

\begin{remark}
In the proof of condition \ref{condition:g1g2} we used the aforementioned
group-entropy relation to translate the problem to the entropy domain. We shall
prove most of the conditions in this manner.
\end{remark}

Observe that the Ingleton inequality has symmetries between subscripts 1
and 2 and between 3 and 4, i.e. if we interchange the subscripts 1
with 2 or 3 with 4, the inequality stays the same. Thus if we prove
some conditions for some $i \in \{1,2\}$ and $j \in \{3,4\}$, we
automatically get conditions for all $(i,j) \in
\{(1,3),(1,4),(2,3),(2,4)\}$. So without loss of generality, we will
just prove conditions for $i\in \{1,3\}$, or $(i,j) \in
\{(1,2),(1,3),(3,1),(3,4)\}$ when these symmetries apply.
\begin{condition}\label{condition:gor1}
$G_i = 1$ or $G$, for some $i$.
\end{condition}
\begin{proof}
For $i = 1$, either would imply $G_1G_2 = G_2G_1$ in
Condition \ref{condition:g1g2}. For $i = 3$, $|G_3|=1$ implies that
the Ingleton inequality becomes $|G_1||G_2||G_{124}|\geq
|G_{12}||G_{14}||G_{24}|$, which clearly follows from $|G_1||G_{124}|\geq
|G_{12}||G_{14}|$ (implied by submodularity of entropy) and $|G_2|\geq
|G_{24}|$.
\end{proof}
\begin{condition}\label{condition:distinct}
$G_i = G_j$ for some distinct $(i,j)$.
\end{condition}
\begin{proof}
(sketch) For $(i,j) = (1,2)$, use $G_1G_2 = G_2G_1$ in Condition
\ref{condition:g1g2}. For $(1,3)$ and $(3,4)$, the argument is similar
to that of the previous condition.
\end{proof}
\begin{condition}\label{condition:g1intxg2}
$G_{12} = 1$.
\end{condition}
\begin{proof}
Realize that Ingleton inequality for entropy vectors can be
rewritten as
\begin{equation}\label{equation:Ingletonsubmod}
r_{13,14}+r_{23,24}+r_{134,234}-r_{123,124} \geq 0,
\end{equation}
where $r_{\alpha,\beta} \triangleq h_\alpha + h_\beta -
h_{\alpha\cap\beta} - h_{\alpha\cup\beta}$ for $\emptyset \neq
\alpha,\beta \subseteq \n$. (e.g., $r_{134,234} = h_{134} + h_{234}
- h_{34} - h_{1234}$.) By submodularity, all $r_{\alpha,\beta} \geq
0$. If $G_{12} = 1$, then $r_{123,124} = 0$ and
(\ref{equation:Ingletonsubmod}) holds.
\end{proof}
\begin{condition}\label{condition:subgroup}
$G_i \leq G_j$ for some distinct $(i,j)$.
\end{condition}
\begin{proof}
(sketch) $(i,j) = (1,2)$ implies $G_1G_2 = G_2G_1$. $(1,3)$ implies
$r_{123,124} = 0$ in (\ref{equation:Ingletonsubmod}). $(3,1)$
implies $r_{123,234} = 0 \Rightarrow r_{123,234} \leq r_{12,24}
\Rightarrow r_{123,124} \leq r_{23,24} \Rightarrow$
(\ref{equation:Ingletonsubmod}) holds. For $(3,4)$, rewrite $h_{13}
= h_{134}$, $h_{23} = h_{234}$, $h_{123} = h_{1234}$, then use
submodularity and non-negativeness of entropy.
\end{proof}
\begin{remark}
Conditions \ref{condition:g1intxg2} and \ref{condition:subgroup}
were first pointed out to us by Prof. M. Aschbacher using group
theoretic techniques. The proof presented above is based on the
submodularity and non-negativity of entropy.
\end{remark}
\begin{remark}
Conditions \ref{condition:abelian}, \ref{condition:g1g2} and
\ref{condition:g1intxg2} are crucial in our searching program, as
they appear in the outer searching loops and can reduce a large
amount of work.
\end{remark}

\section{The Smallest Violation Instance and Its Structure}
Using GAP we found the smallest group that violates Ingleton is $G =
S_5$. There are 60 sets of violating subgroups if we eliminate the
influence of subscript symmetries. Furthermore, these 60 sets of
subgroups are all conjugates of each other. Thus in terms of group
structure, these instances are virtually the same. We list below
some information from GAP about one representative: (the
permutations are written in cycle notation, e.g. $(3,4,5)$ is the
permutation that maps element 3 to 4, element 4 to 5, and element 5 to
3).
\begin{displaymath}
\begin{array}{lll}
G_1 = \langle (3,4,5), (1,2)(4,5) \rangle & \cong S_3 \cong D_6 & |G_1| = 6\\
G_2 = \langle (1,2,3,4,5), (1,4,3,5) \rangle & \cong \z_5 \rtimes
\z_4 & |G_2| = 20\\
G_3 = \langle (2,3), (1,3,4,2) \rangle & \cong D_8 & |G_3| = 8\\
G_4 = \langle (2,4), (1,2,5,4) \rangle & \cong D_8 & |G_4| = 8\\
G_{12} = \langle (1,2)(3,5) \rangle & \cong \z_2 & |G_{12}| = 2\\
G_{13} = \langle (1,2)(3,4) \rangle & \cong \z_2 & |G_{13}| = 2\\
G_{14} = \langle (1,2)(4,5) \rangle & \cong \z_2 & |G_{14}| = 2\\
G_{23} = \langle (1,3,4,2) \rangle & \cong \z_4 & |G_{23}| = 4\\
G_{24} = \langle (1,2,5,4) \rangle & \cong \z_4 & |G_{24}| = 4\\
G_{34} = 1 & & |G_{34}| = 1\\
G_{123} = 1 & & |G_{123}| = 1\\
G_{124} = 1 & & |G_{124}| = 1\\
\end{array}
\end{displaymath}
As $|G_1||G_2||G_{34}||G_{123}||G_{124}| = 120 < 128 =
|G_{12}||G_{13}||G_{14}||G_{23}||G_{24}|$, Ingleton is violated.
Also $G_1$---$G_4$ generate $G = S_5$.

\begin{figure}
\centering
\input{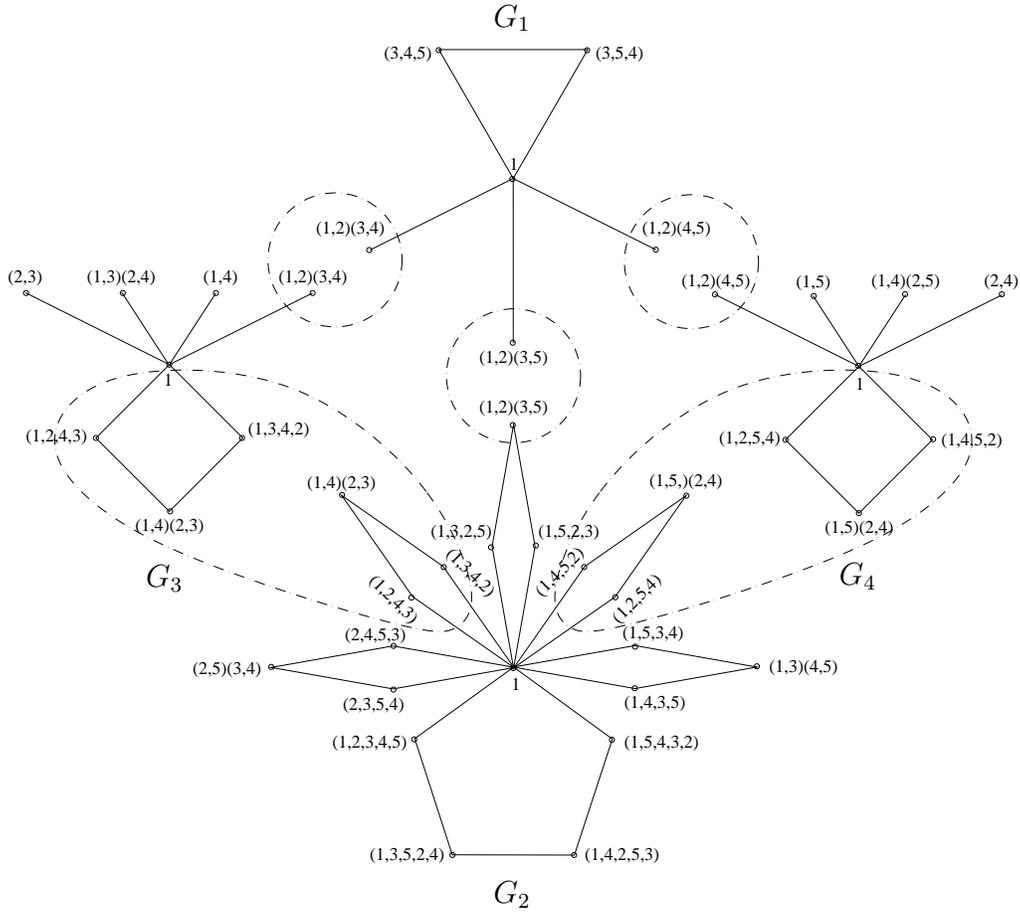}
\caption{The cycle graph of the Ingleton violating subgroups of
$S_5$} \label{figure:flowerstructure}
\end{figure}

To illustrate the structure of these subgroups, we use the group
cycle graph. See Fig.\ref{figure:flowerstructure}, where the
dash-dotted lines denote the pairwise intersections of subgroups
excluding identity.
From the cycle graph we can obtain more structural information which
GAP does not show us directly. First, not only is $G_2$ a semidirect
product of two cyclic groups $\langle (1,2,3,4,5)\rangle \cong \z_5$
and $\langle (1,4,3,5)\rangle \cong \z_4$ (in particular, it's
metacyclic), but also $G_2\setminus\langle (1,2,3,4,5)\rangle
\cup\{1\}$ is the union of subgroups which are all isomorphic to
$\langle (1,4,3,5)\rangle$ (actually they are all conjugates of
$\langle (1,4,3,5)\rangle$) and have trivial pairwise intersections.
(In this case we say $G_2$ has a ``flower'' structure.) Second,
$G_4$ is the conjugate of $G_3$ by $(3,4,5)$ in $G_1$. In
particular, $(1,3,4,2)^{(3,4,5)} = (1,4,5,2) = (1,2,5,4)^{-1}$.

As these subgroups are represented in permutations, it is not easy
either to construct a code from them, or to extend them to a family
of violations. Naturally one may try $S_6$ with similar subgroups,
but unfortunately they do not work. A better way to extract the
structural information and extend the subgroups to a family of
(possible Ingleton-violating) instances, is through the abstract
presentation of groups. It might still be difficult to see concrete
group elements or to prove the structure is successfully extended,
however, we can feed the (extended) presentation to GAP and it might
determine a concrete isomorphic group, which preserves the structure
of violation.

Observe that $|G_{23}| =¡¡|G_{24}| = 4$ contribute most to the $RHS$
of (\ref{equation:Ingletong}), we may try to let the ``petals'' of
$G_2$ (conjugates of $\langle (1,4,3,5)\rangle$) grow while keep
other structures fixed. (This is a little conservative, but it is
the only successful extension according to our GAP trials. For
example, one may try to extend $G_1$ at the same time, but the
structure of $G_3$ and $G_4$ usually collapse.) As $G_2$ plays the
most important role in the violation, we can start from extending
the flower structure of $G_2$. Specifically, we may assume that $G_2
= \langle a,b\rangle$ which has a normal subgroup $N = \langle
a\rangle \cong\z_n$, as well as a subgroup $H = \langle b\rangle
\cong\z_m$, for some generators $a,b$ and integers $m,n$. This gives
us a presentation
\begin{equation}\label{equation:presentationg2}
G_2 = \langle a,b\,|\,a^n=b^m=1,\, a^b=a^s\rangle
\end{equation}
for some $0<s<n$. In order to violate Ingleton as much as possible,
we may wish $n$ to be small while $m$ large. However, the flower
structure of $G_2$ may limit the choices of $n$ and $m$. First of
all, for this presentation to be a semidirect product, we need
$s^m\equiv 1\pmod n$ (see \cite{presGroup}, 5.4). In this case $a,b$
have order $n,m$ respectively, $|G_2| = mn$, $H\cap N = 1$, $s\in
\z_n^\times$ with $|s|\,|\,m$, also $(a^i)^{b^{k}}=a^{is^k}$ for any
integers $i$ and $k$. Moreover, we need $G_2\setminus N \cup\{1\}$
to be the union of groups which are all isomorphic to $H$ with
trivial pairwise intersections.

One possible way to achieve this is to restrict $H^{g_1}\cap H^{g_2}
= 1\ \forall g_1\neq g_2\in N$, as in our original construction.
This is equivalent to $H^g\cap H = 1$, $\forall g\in
N\setminus\{1\}$. If this is the case, there would be $|N|=n$
``petals'' of size $m$ in $G_2$ and the total number of nonidentity
elements would equal $n(m-1) = nm - n = |G_2\setminus N|$, so indeed
the flower structure would be achieved.

Pick two arbitrary nonidentity elements $h_1 = b^l\in H$, $h_2 =
(b^k)^{a^i}\in H^{a^i}$ for some $0<k,l<m$, $0<i<n$. $h_1 =
h_2\Leftrightarrow$ $a^{-i}b^ka^i=b^l \Leftrightarrow
a^{-i}(a^i)^{b^{-k}}b^k=b^l$ $\Leftrightarrow
a^{-i}a^{is^{-k}}=b^{l-k}\Leftrightarrow a^{(s^{-k}-1)i}=b^{l-k}$.
As $H\cap N = 1$, this is equivalent to $a^{(s^{-k}-1)i}=b^{l-k}=1$.
i.e. $l=k$, $n|(s^{-k}-1)i$. To guarantee $H^{a^i}\cap H = 1$ for
any $0<i<n$, we must have $m\leq |s|$. Otherwise we can just choose
$0<k=|s|<m$, then $s^{-k} \equiv 1 \pmod n\Rightarrow n|(s^{-k}-1)i$
for any $i$, and we find a nonidentity element $h_1 = b^k=h_2 =
(b^k)^{a^i}$ in $H^{a^i}\cap H$. So $m\leq |s|$ with $|s|\,|\,m
\Rightarrow m=|s|$. In particular, $m\leq \phi(n) < n$, where
$\phi(n) = |\z_n^\times|$ is Euler's totient function.

For $m$ to be as large as possible, $s$ should be a primitive root
modulo $n$, which makes $m=\phi(n)$. Furthermore, if we choose $n=p$
for some prime $p>2$, then $m = \phi(p) = p-1$ is relatively
``maximized''. (We need $p>2$ for the petals not to collapse.) Also
in this situation if we let $0<k<m=|s|$, $0<i<n=p$, then
$n|(s^{-k}-1)i$ requires $p|i$ or $p|(s^{-k}-1)$. As $p$ is prime,
$p\nmid i$, so $p|(s^{-k}-1)\Rightarrow s^{-k} \equiv 1 \pmod
p\Rightarrow |s|\,|\,k$. But $0<k<|s|$, contradiction. So actually
we have $H^{g}\cap H = 1$, $\forall g\in N$ and the flower structure
is realized.

Now assume $n$, $m$ and $s$ are as above. The next step is to extend
presentation (\ref{equation:presentationg2}) to the whole group $G$
generated by $G_1$---$G_4$. Consider the dihedral groups $G_3$ and
$G_4$. The subgroups of rotations are just $H^{a_3}$ and $H^{a_4}$
respectively, for some $a_3=a^{k_3},a_4=a^{k_4}\in N$. Also $G_3$
and $G_4$ each shares one element of reflection with the dihedral
group $G_1$, while the remaining reflection of $G_1$ is just
$(b^{\frac{p-1}{2}})^{a_1}$ in $G_2$, for some $a_1=a^{k_1}\in N$.
Thus if we can determine the generator of the subgroup of rotations
of $G_1$, then all elements of $G_1$---$G_4$ are determined. In
other words, if we introduce an element $c$ as the generator of
rotations of $G_1$, then all elements from $G_1$---$G_4$ can be
express as products of $a,b,c$ and their inverses. Let's define the
following quantities:
\begin{equation}\label{equation:b1b3b4}
b_1 = (b^{\frac{p-1}{2}})^{a^{k_1}},\quad b_3 = b^{a^{k_3}},\quad
b_4 = b^{a^{k_4}}
\end{equation}
for some integers $k_1,k_3,k_4$. Then we can write
\begin{equation}\label{equation:presentationsubgrps}
G_1 = \langle c,b_1\rangle,\quad G_2 = \langle a,b\rangle,\quad G_3
= \langle b_1c^2,b_3\rangle,\quad G_4 = \langle
b_1c,b_4\rangle,\quad G = \langle a,b,c\rangle.
\end{equation}
As $G_1\cong D_6$, we should have the relation $c^3=(cb_1)^2=1$. For
$G_3$ and $G_4$ to be dihedral groups, we need
$(b_3b_1c^2)^2=(b_4b_1c)^2=1$. We may use GAP to determine a
concrete group with these relations, but there are still too many
parameters to choose and we do not know which ones may yield the
correct structure.

Observe in the original violation, the structure $G_4 =
G_3^{(3,4,5)}$ with generators $(1,3,4,2)^{(3,4,5)} =
(1,2,5,4)^{-1}$ is not utilized yet. If we let $a=(1,2,3,4,5)$,
$b=(1,4,3,5)$, $c = (3,4,5)$, $b_3 = (1,3,4,2)$, $b_4 = (1,2,5,4)$
in the original construction, then the relation above translates to
$b_3^c = b_4^{-1}$. We claim this relation for our presentation
automatically makes $(b_3b_1c^2)^2=(b_4b_1c)^2=1$ if and only if
$k_3 - k_1 \equiv k_1 - k_4 \pmod p$: as $|b_1| = 2$,
$c^3=(cb_1)^2=1\Rightarrow cb_1 =b_1c^2$, $(b_3b_1c^2)^2 =
b_3b_1c^{-1}b_3cb_1 = b_3b_1b_4^{-1}b_1$ by the new relation.
Similarly $(b_4b_1c)^2 = b_4b_1b_3^{-1}b_1 =
((b_3b_1c^2)^{-2})^{b_1}$, so $(b_3b_1c^2)^2=1 \Leftrightarrow
(b_4b_1c)^2=1$. Plugging in (\ref{equation:b1b3b4}) and using
$(a^i)^{b^{k}}=a^{is^k}$ we have
\begin{displaymath}
(b_3b_1c^2)^2 = a^{[(k_3-k_1)+(k_1-k_4)s^{(p-1)/2}](s^{-1}-1)}.
\end{displaymath}
Since $s$ is a primitive root modulo $p$, $|s^{(p-1)/2}|=2$. As
$\z_p^\times$ is cyclic of an even order $p-1$, it is clear that
there is a unique element of order $2$. Also $|(p-1)| = 2$ as
$(p-1)^2 \equiv 1 \pmod p$, so $s^{(p-1)/2}=p-1$, and
\begin{displaymath}
(b_3b_1c^2)^2 = a^{[(k_3-k_1)-(k_1-k_4)](s^{-1}-1)}.
\end{displaymath}
Now $p\nmid(s^{-1}-1)$ as $|s|=p-1$, then $(b_3b_1c^2)^2 = 1$ if and
only if $p|[(k_3-k_1)-(k_1-k_4)]$ if and only if $k_3 - k_1 \equiv
k_1 - k_4 \pmod p$. This condition gives us a smaller set of
parameters as well as a simpler presentation, while maintains all
the structures of the subgroups. (Actually once $k_3 - k_1 \equiv
k_1 - k_4 \pmod p$ is satisfied, it is very easy to use GAP to
produce the desired structures, even with arbitrary $k_1$ and
$k_3$.)

In sum, our analysis gives us the following presentation:
\begin{equation}\label{equation:presentationg}
G = \langle a,b,c\,|\,a^p=b^{p-1}=c^3=1,\, a^b=a^s,\,
(cb_1)^2=b_3^cb_4=1\rangle
\end{equation}
where $p$ is an odd prime, $s$ is a primitive root modulo $p$, $k_3
- k_1 \equiv k_1 - k_4 \pmod p$. If our extension of the subgroup
structures succeeds, then the orders of subgroups and intersections
would be: $|G_1| = 6$, $|G_2| = p(p-1)$, $|G_3| = |G_4| = 2(p-1)$,
$|G_{12}| = |G_{13}| = |G_{14}| = 2$, $|G_{23}| = |G_{24}| = p-1$,
$|G_{34}| = |G_{123}| = |G_{124}| =1$. $LHS$ of
(\ref{equation:Ingletong}) $=6p(p-1)$ while $RHS = 8(p-1)^2$. So for
$p\geq5$, Ingleton should be violated.

\section{Explicit Violation Construction with $PGL(2,p)$}
Plugging the above presentation into GAP with different $p$'s and
other parameters, we get a series of groups. However, when $p$ is
large, GAP usually runs out of memory for some (even simple)
operations. According to our computation, for $p = 5,7,\dots,23$ GAP
determined that they are all finite groups and all violate Ingleton.
Among these groups GAP determined their isomorphism types up to $p =
19$, most of which are semidirect products $PSL(2,p)\rtimes\z_2$. As
$PGL(2,p)$'s are also semidirect products of $PSL(2,p)$ and $\z_2$,
and $PGL(2,5) \cong S_5$, we guess the isomorphism type for these
groups might just be $PGL(2,p)$. This conjecture is verified by GAP
up to $p = 11$.

Although $PGL(2,p)$'s are relatively easy groups of matrices, GAP
uses isomorphic permutation groups to represent them. This makes it
difficult to recognize the corresponding matrices of the output
subgroups. However, with presentation (\ref{equation:presentationg})
we may explicitly identify the generators in $PGL(2,p)$ and check
their relations, then use (\ref{equation:presentationsubgrps}) to
construct the subgroups.

Let $p$ be an odd prime. For $A\in GL(2,p)$, let $\A$ denote the
left coset of $A$ in $GL(2,p)$ with respect to $V = \{\alpha
I:\alpha \in\f_p^\times\}$. Thus $\A = \B$ if and only if each entry
of $A$ is a nonzero constant multiple of the corresponding entry of
$B$. We denote the elements of $\f_p$ by ordinary integers, but the
addition and multiplication, as well as equality, are modulo $p$.
Furthermore, $-k$ and $k^{-1}$ denotes the additive and
multiplicative inverses of $k$ in $\f_p$ respectively. This would
not cause any confusion as we only use elements from $\f_p$ in the
entries of matrices.

Consider the following matrices in $GL(2,p)$:
\begin{equation}\label{equation:ABC}
A = \left[ \begin{array}{cc} 1 & 0 \\
1 & 1 \end{array}\right],\quad B = \left[ \begin{array}{cc} 1 & 0 \\
0 & t \end{array}\right],\quad C = \left[
\begin{array}{cc} 1 & \frac{p-1}{2} \\ 2 & 0 \end{array}\right]
\end{equation}
where $t$ is a primitive root modulo $p$, i.e. a generator of
$\f_p^\times$. Our guess is $\A,\B,\C$ corresponds to the generators
$a,b,c$ in (\ref{equation:presentationg}) respectively. The powers
of these matrices are:
\begin{displaymath}
A^k = \left[ \begin{array}{cc} 1 & 0 \\
k & 1 \end{array}\right],\quad
B^k = \left[ \begin{array}{cc} 1 & 0 \\
0 & t^k \end{array}\right],\quad
C^2 = \left[\begin{array}{cc} 0 &
\frac{p-1}{2} \\ 2 & -1 \end{array}\right],\quad
C^3=\left[\begin{array}{cc} p-1 & 0 \\ 0 & p-1\end{array}\right]
\end{displaymath}
for any integer $k$. Thus $\A^p = \I$, $\B^{p-1} = \I$, $\C^3 = \I$
and $|\A| = p$, $|\B| = p-1$, $|\C| = 3$. Also
\begin{displaymath}
A^B = B^{-1}AB = \left[ \begin{array}{cc} 1 & 0 \\
0 & t^{-1} \end{array}\right]\left[ \begin{array}{cc} 1 & 0 \\
1 & 1 \end{array}\right]\left[ \begin{array}{cc} 1 & 0 \\
0 & t \end{array}\right] = \left[ \begin{array}{cc} 1 & 0 \\
t^{-1} & 1 \end{array}\right] = A^s,
\end{displaymath}
where $s = t^{-1}$ is also a primitive root modulo $p$. So $\A^{\B}
= \A^s$. Next we let
\begin{displaymath}
B_1 = (B^{\frac{p-1}{2}})^{A^{k_1}}
 = A^{-k_1}B^{\frac{p-1}{2}}A^{k_1} = \left[ \begin{array}{cc} 1 & 0 \\
-k_1 & 1 \end{array}\right]\left[ \begin{array}{cc} 1 & 0 \\
0 & -1 \end{array}\right]\left[ \begin{array}{cc} 1 & 0 \\
k_1 & 1 \end{array}\right] = \left[ \begin{array}{cc} 1 & 0 \\
-2k_1 & -1 \end{array}\right],
\end{displaymath}
where we calculated $t^{\frac{p-1}{2}} = -1$ as it is the unique
element of order $2$ in $\f_p^\times$. Now check
\begin{displaymath}
CB_1 = \left[\begin{array}{cc} 1 & \frac{p-1}{2} \\ 2 & 0
\end{array}\right] \left[ \begin{array}{cc} 1 & 0 \\ -2k_1 & -1
\end{array}\right] = \left[ \begin{array}{cc} 1+k_1 & \frac{p+1}{2} \\ 2 &
0\end{array}\right],
\end{displaymath}
\begin{displaymath}
(CB_1)^2 = \left[ \begin{array}{cc} 1+k_1 & \frac{p+1}{2} \\ 2 &
0\end{array}\right]^2 = \left[ \begin{array}{cc} (1+k_1)^2+1 &
(1+k_1)\frac{p+1}{2} \\ 2(1+k_1) & 1\end{array}\right].
\end{displaymath}
Thus if we want $(\C\Ba)^2 = \I$, $k_1$ must be $-1$. In this case
\begin{displaymath}
\Ba = \overline{\left[ \begin{array}{cc} 1 & 0 \\2 & -1
\end{array}\right]},\quad
\C\Ba = \overline{\left[\begin{array}{cc} 0 & \frac{p+1}{2} \\ 2 &
0\end{array}\right]} = \overline{\left[\begin{array}{cc} 0 &
\frac{p-1}{2} \\ -2 & 0\end{array}\right]},\quad (\C\Ba)^2 = \I.
\end{displaymath}
Next we calculate:
\begin{displaymath}
B^{A^k} = A^{-k}BA^{k} = \left[ \begin{array}{cc} 1 & 0 \\
-k & 1 \end{array}\right]\left[ \begin{array}{cc} 1 & 0 \\
0 & t \end{array}\right]\left[ \begin{array}{cc} 1 & 0 \\
k & 1 \end{array}\right] = \left[ \begin{array}{cc} 1 & 0 \\
k(t-1) & t \end{array}\right].
\end{displaymath}
Let $B_3=B^{A^{k_3}}$, $B_4=B^{A^{k_4}}$. As $k_1 = -1$, $k_3 - k_1
= k_1 - k_4$, we have $k_4=-2-k_3$.
\begin{eqnarray*}
B_3CB_4 &=& \left[ \begin{array}{cc} 1 & 0 \\
k_3(t-1) & t \end{array}\right] \left[\begin{array}{cc} 1 &
\frac{p-1}{2} \\ 2 & 0 \end{array}\right] \left[ \begin{array}{cc} 1 & 0 \\
k_4(t-1) & t \end{array}\right]\\
&=& \left[ \begin{array}{cc} 1 & \frac{p-1}{2} \\ k_3(t-1)+2t &
k_3(t-1)\frac{p-1}{2}\end{array}\right] \left[ \begin{array}{cc} 1 &
0 \\ k_4(t-1) & t \end{array}\right],
\end{eqnarray*}
whose $(2,2)$-entry is $k_3(t-1)\frac{p-1}{2}t$. If we want
$(\Bc)^{\C}\cdot\Bd=\I \Leftrightarrow \Bc\C\Bd=\C$, $k_3$ must be
$0$ as the $(2,2)$-entry of $C$ is $0$ and all $t-1,\frac{p-1}{2},t$
are nonzero. So $k_4=-2-k_3=-2$,
\begin{displaymath}
\Bc = \overline{\left[ \begin{array}{cc} 1 & 0 \\
0 & t \end{array}\right]} = \B,\quad
\Bd =\overline{\left[ \begin{array}{cc} 1 & 0 \\
2(1-t) & t \end{array}\right]},
\end{displaymath}
\begin{displaymath}
\Bc\C\Bd = \overline{\left[ \begin{array}{cc} 1 & \frac{p-1}{2} \\
2t & 0
\end{array}\right] \left[ \begin{array}{cc} 1 & 0 \\ 2(1-t) & t
\end{array}\right]} =
\overline{\left[ \begin{array}{cc} t & \frac{p-1}{2}t \\ 2t & 0
\end{array}\right]} = \C.
\end{displaymath}

So far for $\A,\B,\C$ we have verified all the relations in
(\ref{equation:presentationg}). We can also prove that they are
actually a set of generators for $PGL(2,p)$. Observe that each
matrix in $GL(2,p)$ can be written as a product of the following
elementary matrices:
\begin{displaymath}
\left[ \begin{array}{cc} 1 & 0 \\
k & 1 \end{array}\right],\quad \left[ \begin{array}{cc} 1 & k \\
0 & 1 \end{array}\right],\quad \left[ \begin{array}{cc} 1 & 0 \\
0 & t^k \end{array}\right],\quad \left[ \begin{array}{cc} t^k & 0 \\
0 & 1 \end{array}\right]
\end{displaymath}
which are generated by $A,A^T,B,t^{-1}B$. So $PGL(2,p)$ is generated
by $\A,\overline{A^T},\B$. Now as $t^l = -2$ for some integer $l$,
$t^{-l} = (-2)^{-1} = \frac{p-1}{2}$. We have
\begin{eqnarray*}
B^{-l}A^{-2}CB^l &=& \left[ \begin{array}{cc} 1 & 0 \\
0 & \frac{p-1}{2} \end{array}\right] \left[\begin{array}{cc} 1 & 0
\\ -2 & 1 \end{array}\right] \left[\begin{array}{cc} 1 &
\frac{p-1}{2} \\ 2 & 0 \end{array}\right] \left[ \begin{array}{cc} 1
& 0 \\ 0 & -2 \end{array}\right]\\
&=& \left[\begin{array}{cc} 1 & 0 \\ 1 & \frac{p-1}{2}
\end{array}\right] \left[\begin{array}{cc} 1 &
1 \\ 2 & 0 \end{array}\right] = \left[\begin{array}{cc} 1 & 1 \\ 0 &
1 \end{array}\right] = A^T,
\end{eqnarray*}
Thus $\A,\B,\C$ also generate $PGL(2,p)$. So if we set $s = t^{-1}$,
$k_1 = -1$, $k_3 =0$, $k_4 = -2$, then $\A,\B,\C$ corresponds to the
generators in (\ref{equation:presentationg}).
\begin{remark}
Note that we have not proved that (\ref{equation:presentationg}) is
a presentation of $PGL(2,p)$. In order to do that, one must show
that for any group generated by $a,b,c$ while satisfying the
relations in (\ref{equation:presentationg}), the order must be no
more than $|PGL(2,p)| = (p-1)p(p+1)$. This is not proved yet.
However, identifying possible corresponding generators still gives
us a way to explicitly construct the subgroups to violate Ingleton.
\end{remark}

Now we can write out the subgroups in $PGL(2,p)$ corresponding to
subgroups in (\ref{equation:presentationsubgrps}).

$G_1 = \langle\C,\Ba\rangle$. Note that $|\C|=3$, $|\Ba|=2$,
$(\C\Ba)^2 = \I \Leftrightarrow \C\Ba = \Ba(\C)^2$, so $G_1$ has at
most $6$ elements $\{(\Ba)^i(\C)^j: 0\leq i <2,\; 0\leq j < 3\}$.
Calculating these elements we can see $|G_1|=6$ exactly:
\begin{displaymath}
G_1 = \left\{\I,\quad
\overline{\left[\begin{array}{cc} 1 & \frac{p-1}{2} \\ 2 & 0
\end{array}\right]},\quad
\overline{\left[\begin{array}{cc} 0 & \frac{p-1}{2} \\ 2 & -1
\end{array}\right]},\quad
\overline{\left[ \begin{array}{cc} 1 & 0 \\2 & -1
\end{array}\right]},\quad
\overline{\left[ \begin{array}{cc} 1 & \frac{p-1}{2} \\0 & -1
\end{array}\right]},\quad
\overline{\left[\begin{array}{cc} 0 & \frac{p-1}{2} \\ -2 &
0\end{array}\right]} \right\}.
\end{displaymath}

$G_2 = \langle\A,\B\rangle$. We claim that $G_2$ is just the
subgroup of lower triangular matrices in $GL(2,p)$ modulo $V$, i.e.
\begin{displaymath}
G_2 = \left\{\overline{\left[\begin{array}{cc} 1 & 0 \\ \alpha &
\beta\end{array}\right]} :\alpha\in\f_p,\
\beta\in\f_p^\times\right\}.
\end{displaymath}
As $A,B$ are lower triangular, any element in $G_2$ is a lower
triangular matrix modulo $V$. On the other hand, $\forall
\alpha\in\f_p,\;\beta\in\f_p^\times$, $\beta = t^l$ for some integer
$l$. So
\begin{displaymath}
\left[\begin{array}{cc} 1 & 0 \\ \alpha & \beta\end{array}\right] =
A^{\alpha}B^l \Rightarrow \overline{\left[\begin{array}{cc} 1 & 0 \\
\alpha & \beta\end{array}\right]} = \A^{\alpha}\B^l \in G_2.
\end{displaymath}
Therefore $|G_2| = p(p-1)$.

$G_3 = \langle\Ba(\C)^2,\Bc\rangle = \langle\C\Ba,\Bc\rangle$. Note
that $|\C\Ba|=2$, $|\Bc|=|\B|=p-1$, also
\begin{displaymath}
\Bc\cdot\C\Ba = \overline{\left[\begin{array}{cc} 1 & 0 \\ 0 & t
\end{array}\right]\left[\begin{array}{cc} 0 &
\frac{p-1}{2} \\ -2 & 0\end{array}\right]} =
\overline{\left[\begin{array}{cc} 0 & \frac{p-1}{2} \\
 -2t & 0\end{array}\right]} =
\overline{\left[\begin{array}{cc} 0 & \frac{p-1}{2}t^{-1} \\
 -2 & 0\end{array}\right]} = \C\Ba(\Bc)^{-1},
\end{displaymath}
so $G_3$ has at most $2(p-1)$ elements $\{(\C\Ba)^i(\Bc)^j: 0\leq i
<2,\; 0\leq j < p-1\}$. Calculating these elements we can see
$|G_3|=2(p-1)$ exactly:
\begin{displaymath}
G_3 = \left\{(\Bc)^k = \overline{\left[\begin{array}{cc} 1 & 0 \\ 0
& t^k \end{array}\right]},\quad \C\Ba(\Bc)^k =
\overline{\left[\begin{array}{cc} 0 & \frac{p-1}{2}t^k \\ -2 &
0\end{array}\right]} :\ 0\leq k < p-1\right\}.
\end{displaymath}

$G_4 = \langle\Ba\C,\Bd\rangle$. Note that
\begin{displaymath}
\Ba\C = \overline{\left[\begin{array}{cc} 1 & \frac{p-1}{2} \\ 0 &
-1\end{array}\right]},\quad
(\Bd)^k = \overline{\left[ \begin{array}{cc} 1 & 0 \\
2(1-t^k) & t^k \end{array}\right]},
\end{displaymath}
so $|\Ba\C|=2$, $|\Bd|=p-1$. Also
\begin{eqnarray*}
\Bd\cdot\Ba\C &=& \overline{\left[\begin{array}{cc} 1 &
\frac{p-1}{2}\\ 2(1-t) & -1 \end{array}\right]} =
\overline{\left[\begin{array}{cc} t^{-1} & \frac{p-1}{2}t^{-1} \\
2(t^{-1}-1) & -t^{-1}\end{array}\right]} \\
&=& \overline{\left[\begin{array}{cc} 1 & \frac{p-1}{2} \\ 0 &
-1\end{array}\right]\left[ \begin{array}{cc} 1 & 0 \\
2(1-t^{-1}) & t^{-1} \end{array}\right]} = \Ba\C(\Bd)^{-1},
\end{eqnarray*}
so $G_4$ has at most $2(p-1)$ elements $\{(\Ba\C)^i(\Bd)^j: 0\leq i
<2,\; 0\leq j < p-1\}$. Calculating these elements we can see
$|G_4|=2(p-1)$ exactly:
\begin{displaymath}
G_4 = \left\{(\Bd)^k = \overline{\left[\begin{array}{cc} 1 & 0 \\
2(1-t^k) & t^k \end{array}\right]},\quad \Ba\C(\Bd)^k =
\overline{\left[\begin{array}{cc} 1 & \frac{p-1}{2}\\ 2(1-t^{-k}) &
-1 \end{array}\right]} :\ 0\leq k < p-1\right\}.
\end{displaymath}

With all four subgroups explicitly written, we can easily write down
the intersections:
\begin{displaymath}
G_{12} = \left\{\I,\quad \overline{\left[ \begin{array}{cc} 1 & 0
\\2 & -1\end{array}\right]}\right\},\quad |G_{12}|=2.
\end{displaymath}
\begin{displaymath}
G_{13} = \left\{\I,\quad \overline{\left[\begin{array}{cc} 0 &
\frac{p-1}{2} \\ -2 & 0\end{array}\right]}\right\},\quad |G_{13}|=2.
\end{displaymath}
\begin{displaymath}
G_{14} = \left\{\I,\quad \overline{\left[ \begin{array}{cc} 1 &
\frac{p-1}{2} \\0 & -1 \end{array}\right]}\right\},\quad |G_{14}|=2.
\end{displaymath}
\begin{displaymath}
G_{23} = \left\{\overline{\left[\begin{array}{cc} 1 & 0 \\
0 & t^k \end{array}\right]} :\ 0\leq k < p-1\right\},\quad
|G_{23}|=p-1.
\end{displaymath}
\begin{displaymath}
G_{24} = \left\{\overline{\left[\begin{array}{cc} 1 & 0 \\
2(1-t^k) & t^k \end{array}\right]} :\ 0\leq k < p-1\right\},\quad
|G_{24}|=p-1.
\end{displaymath}
\begin{displaymath}
G_{34} = G_{123} = G_{124} = 1.
\end{displaymath}
So in (\ref{equation:Ingletong}), indeed $LHS =
|G_1||G_2||G_{34}||G_{123}||G_{124}| = 6p(p-1)$, $RHS =
|G_{12}||G_{13}||G_{14}||G_{23}||G_{24}| = 8(p-1)^2$, $LHS - RHS =
2(p-1)(4-p)$. Thus Ingleton is violated when $p\geq 5$, and the subgroup
structures of $S_5$ are exactly reproduced.

\section{Conclusion}
Using a refined search we found the smallest group to violate the
Ingleton inequality to be the 120 element group $S_5$. Investigating
the detailed structure of the subgroups allowed us to determine that
this is an instance of the Ingleton-violating family of groups
$PGL(2,p)$ for primes $p\geq 5$. We have begun investigating
$PGL(2,p^q)$ groups and conjecture that they violate Ingleton for
large enough $p$ and $q$. Computer search verifies that $PGL(2,2^2)$
does not violate Ingleton, whereas $PGL(2,2^3)$ and $PGL(2,3^2)$ do.
Finally, investigating the use of these groups to construct network
codes more powerful than linear ones may be a fruitful direction for
future work.

\appendices


\section*{Acknowledgment}
The authors would like to thank Michael Aschbacher and Amin
Shokrollahi for very helpful discussions on the conditions
and on expanding the group structures.



\bibliographystyle{IEEEtran}
\bibliography{IEEEabrv,IngletonVioDraft}
%
%
%

\end{document}